\documentclass[11pt]{article}
\pdfoutput=1
\usepackage{fullpage}
\usepackage{xcolor}
\usepackage{cite}
\usepackage{graphicx}
\usepackage{amsmath}
\usepackage{amssymb}
\usepackage{subcaption}
\usepackage[utf8x]{inputenc} 
\usepackage{todonotes}
\title{}
\date{}

\def\beq{\begin{equation}}
\def\eeq{\end{equation}}
\allowdisplaybreaks
\begin{document}
\bibliographystyle{utphys}

% Commands
\newcommand{\be}{\begin{equation}}
\newcommand{\ee}{\end{equation}}
\newcommand\n[1]{\textcolor{red}{(#1)}} %in-text notes
\newcommand{\diff}{\mathop{}\!\mathrm{d}}
\newcommand{\lb}{\left}
\newcommand{\rb}{\right}
\newcommand{\f}{\frac}
\newcommand{\pd}{\partial}
\newcommand{\tr}{\text{tr}}
\newcommand{\fdiff}{\mathcal{D}}
\newcommand{\im}{\text{im}}
\let\caron\v
\renewcommand{\v}{\mathbf}
\newcommand{\T}{\tensor}
\newcommand{\R}{\mathbb{R}}
\newcommand{\C}{\mathbb{C}}
\newcommand{\Z}{\mathbb{Z}}
\newcommand{\msbar}{\ensuremath{\overline{\text{MS}}}}
\newcommand{\DIS}{\ensuremath{\text{DIS}}}
\newcommand{\abar}{\ensuremath{\bar{\alpha}_S}}
\newcommand{\bb}{\ensuremath{\bar{\beta}_0}}
\newcommand{\rc}{\ensuremath{r_{\text{cut}}}}
\newcommand{\Nd}{\ensuremath{N_{\text{d.o.f.}}}}
\newcommand{\red}[1]{{\color{red} #1}}
\newcommand{\blue}{\color{blue}}
\newcommand{\black}{\color{black}}
\setlength{\parindent}{0pt}

  \newcommand\snNote[1]{
 \todo[backgroundcolor=red!10!white,fancyline,
 bordercolor=white]{SN: #1}}

\titlepage
\begin{flushright}
QMUL-PH-21-58\\
\end{flushright}

\vspace*{0.5cm}

\begin{center}
{\bf \Large Alternative formulations of the twistor double copy}

\vspace*{1cm} 
\textsc{Erick Chac\'{o}n\footnote{e.c.chaconramirez@qmul.ac.uk},
Silvia Nagy\footnote{s.nagy@qmul.ac.uk},
  and Chris D. White\footnote{christopher.white@qmul.ac.uk}} \\

\vspace*{0.5cm} Centre for Theoretical Physics, Department of Physics
and Astronomy, \\ Queen Mary University of London, Mile End Road,
London E1 4NS, UK\\

\end{center}

\vspace*{0.5cm}

\begin{abstract}
The classical double copy relating exact solutions of biadjoint
scalar, gauge and gravity theories continues to receive widespread
attention. Recently, a derivation of the exact classical double copy
was presented, using ideas from twistor theory, in which spacetime
fields are mapped to \u{C}ech cohomology classes in twistor space. A
puzzle remains, however, in how to interpret the twistor double copy,
in that it relies on somehow picking special representatives of each
cohomology class. In this paper, we provide two alternative
formulations of the twistor double copy using the more widely-used
language of Dolbeault cohomology. The first amounts to a rewriting of
the \u{C}ech approach, whereas the second uses known techniques for
discussing spacetime fields in Euclidean signature. The latter
approach indeed allows us to identify special cohomology
representatives, suggesting that further application of twistor
methods in exploring the remit of the double copy may be fruitful.
\end{abstract}

\vspace*{0.5cm}

\section{Introduction}
\label{sec:intro}

The study of (quantum) field theories in recent years has been
characterised by a relentless search for common underlying
structures. An example of this endeavour is the {\it double copy}, a
set of ideas for relating various quantities in a number of different
theories. Inspired by previous work in string
theory~\cite{Kawai:1985xq}, the double copy was first formulated for
scattering amplitudes in gauge and gravity
theories~\cite{Bern:2010ue,Bern:2010yg}, both with and without
supersymmetry. It was subsequently extended to exact classical
solutions in ref.~\cite{Monteiro:2014cda}, which focused on the
special -- but infinite -- family of {\it Kerr-Schild} solutions in
gravity. Follow-up work (see e.g.
refs.~\cite{Luna:2015paa,Ridgway:2015fdl,Bahjat-Abbas:2017htu,Berman:2018hwd,Carrillo-Gonzalez:2017iyj,CarrilloGonzalez:2019gof,Bah:2019sda,Alkac:2021seh,Bahjat-Abbas:2020cyb,Alfonsi:2020lub})
has attempted to see whether this family of solutions can be extended,
and the development of different techniques is also useful in this
regard. Reference~\cite{Luna:2018dpt} (see also
refs.~\cite{Sabharwal:2019ngs,Alawadhi:2020jrv,Godazgar:2020zbv})
presented an alternative exact classical double copy, that uses the
spinorial formalism of General Relativity and related field theories,
and which is known as the {\it Weyl double copy}. This is more general
than the Kerr-Schild approach of ref.~\cite{Monteiro:2014cda}, and
includes the latter as a special case. To date it remains the most
general exact classical double copy procedure, although alternative
formalisms offer complementary
insights~\cite{Elor:2020nqe,Farnsworth:2021wvs,Anastasiou:2014qba,LopesCardoso:2018xes,Anastasiou:2018rdx,Luna:2020adi,Borsten:2020xbt,Borsten:2020zgj,Borsten:2021hua,Campiglia:2021srh},
and it is also known how to double-copy classical solutions
order-by-order in the coupling constants of given physical theories,
at the price of giving up exactness (see
e.g. refs.~\cite{Luna:2016due,Luna:2016hge,Luna:2017dtq,Goldberger:2017frp,Goldberger:2017vcg,Goldberger:2017ogt,Goldberger:2019xef,Goldberger:2016iau,Prabhu:2020avf}). This
may offer new calculational tools for astrophysical observables,
including those related to gravitational waves. However, it is also
important to probe the origins of the double copy, given that a fully
nonperturbative understanding of its scope and applicability is still
missing\footnote{For recent proofs of the double copy in a
perturbative field theory context, see
refs.~\cite{Borsten:2021hua,Borsten:2021rmh}.}. \\

Recently, refs.~\cite{White:2020sfn,Chacon:2021wbr} provided a
derivation of the Weyl double copy using twistor
theory~\cite{Penrose:1967wn,Penrose:1972ia,Penrose:1968me,Woodhouse:1985id}
(see
e.g. refs.~\cite{Penrose:1987uia,Penrose:1986ca,Huggett:1986fs,Adamo:2017qyl,Wolf_2010}
for pedagogical reviews of this subject, and
refs.~\cite{Farnsworth:2021wvs,Chacon:2021hfe} for related work on
twistor approaches to the double copy). Basic ideas from the latter
include that points in our own spacetime are mapped non-locally to
geometric objects in an abstract twistor space ${\mathbb T}$, and vice
versa. Furthermore, physical fields in spacetime map to cohomological
data in twistor space. In more pedestrian terms, one may write
solutions of the field equation for massless free spacetime fields as
a certain contour integral in twistor space known as the {\it Penrose
  transform}. The integrand contains a holomorphic function of twistor
variables, which is defined up to contributions that vanish upon
performing the integral. The freedom to redefine twistor functions in
this manner is expressed by saying that they are cohomology classes
(elements of a cohomology group), and in the traditional twistor
theory approach pioneered by
refs.~\cite{Penrose:1967wn,Penrose:1972ia,Penrose:1968me}, these are
{\it sheaf cohomology groups}, which can be suitably approximated by
{\it \u{C}ech cohomology groups}. \\

The \u{C}ech approach was used by
refs.~\cite{White:2020sfn,Chacon:2021wbr} to derive the Weyl double
copy, which led to an interesting puzzle. The spacetime relationship
embodied by the Weyl double copy turns into a simple product of
functions in the integrand of the Penrose transform in twistor
space. As remarked above, however, these are not actually functions,
but representatives of cohomology classes, which are meant to be
subjectable to the above-mentioned redefinitions. Any non-linear
relationship is incompatible with first performing such redefinitions,
and thus it seems that the twistor approach demands certain
``special'' representatives of each cohomology class be chosen, with
no useful guidance of how to make such a choice. All that is needed to
derive the Weyl double copy in position space is simply to find
suitable representatives in twistor space that do the right job. But
it would be nice to know if the double copy can be given a more
genuinely twistorial interpretation, by fixing a procedure for
choosing appropriate representatives.\\

Another potential issue with refs.~\cite{White:2020sfn,Chacon:2021wbr}
is that the \u{C}ech approach is not so widely used in contemporary
works on twistor approaches to field theory. Instead, it is more
common to use the language of differential forms, where the
ambiguities inherent in the Penrose transform can be characterised by
{\it Dolbeault cohomology}~\cite{WoodhouseTN,bams/1183544081}. That
this is equivalent to the \u{C}ech approach follows from known
isomorphisms between \u{C}ech and Dolbeault cohomology groups. Thus,
if the double copy has a genuinely twistorial expression, then it must
be possible to describe it using the Dolbeault language. Preliminary
and very useful comments in this regard were made in
ref.~\cite{Adamo:2021dfg}, which presented a classical double copy
defined at asymptotic infinity in spacetime, and showed that it could
be used to constrain Dolbeault representatives in the twistor
formalism (see ref.~\cite{MasonTN} for earlier related work). Our aim
in this paper is to explore the relationship between the Dolbeault and
\u{C}ech approaches in more detail, and also to go beyond the purely
radiative spacetimes considered in ref.~\cite{Adamo:2021dfg}. We will
present two different incarnations of the Dolbeault double copy. The
first is ultimately a rewriting of the \u{C}ech approach, using a
known approach for turning representatives of \u{C}ech cohomology
groups into Dolbeault representatives. We will see that a product
structure in twistor space indeed emerges in the Dolbeault framework,
which is ultimately not surprising given that this is essentially
inherited from the \u{C}ech double copy. Furthermore, this first
technique for constructing a Dolbeault double copy will suffer from
the same inherent ambiguities as the \u{C}ech approach, namely that it
is not clear what the recipe is for picking out a special
representative of each cohomology class. Motivated by this puzzle, we
will then present a second Dolbeault double copy, which uses known
techniques for writing Dolbeault representatives associated with
spacetime fields in Euclidean signature. We will argue that the
spacetime double copy is again associated with a certain product of
functions in twistor space. In this case, however, special
representatives of each cohomology class are indeed picked out: they
are the {\it harmonic} representatives, which are uniquely defined for
each spacetime field. We hope that our results provide further
motivation for the use of twistor methods in understanding the
classical double copy. They may also prove useful in relating the
classical double copy with the original BCJ double copy for scattering
amplitudes, given that twistor methods have appeared naturally in the
study of latter (see
e.g. refs.~\cite{Mason:2013sva,Geyer:2014fka,Casali:2015vta,Casali:2016atr,Adamo:2017sze}). \\

The structure of our paper is as follows. In section~\ref{sec:review},
we review the twistor double copy of
refs.~\cite{White:2020sfn,Chacon:2021wbr}, using the \u{C}ech
formalism, and also relevant aspects of differential forms and
Dolbeault cohomology needed for what follows. In
section~\ref{sec:Dolbeault}, we provide a first example of the twistor
double copy in the Dolbeault language, and demonstrate its close
relation to the \u{C}ech approach. In section~\ref{sec:Euclidean}, we
provide a second incarnation, and argue that it allows us to identify
special representatives of each cohomology class. We discuss the
implications of our results in section~\ref{sec:discuss}.

\section{Review of necessary concepts}
\label{sec:review}

In this section, we will review those details of twistor theory that
are needed for what follows, including relevant aspects of \u{C}ech
and Dolbeault cohomology. We will also describe the twistor double
copy of refs.~\cite{White:2020sfn,Chacon:2021wbr}, which was
formulated in the \u{C}ech language. All of these ideas rely on the
spinorial formalism of field theory, in which any spacetime tensor
field\footnote{Throughout, we will use lower-case Latin letters for
spacetime indices, upper-case Latin letters for spinor indices, and
Greek letters for the twistor indices to be defined in what follows.}
can be converted to a multi-index spinor upon contracting with {\it
  Infeld-van-der-Waerden} symbols $\{\sigma^a_{AA'}\}$, defined in a
suitable basis\footnote{A common choice results in the identity matrix
for $\sigma^0_{AA'}$, and the Pauli matrices for $\sigma^i_{AA'}$.}
e.g.
\begin{equation}
  V_{AA'}=V_a \sigma^a_{AA'}.
  \label{spinorial}
\end{equation}
Spinor indices $A$ and $A'$ run from 0 to 1, and can be raised and
lowered with the 2-dimensional Levi-Civita symbols $\epsilon^{AB}$,
$\epsilon^{A'B'}$ etc. The spinorial formalism makes many nice
properties of field theory manifest. In particular, any multi-index
spinor can be decomposed into sums of fully symmetric spinors
multiplied by Levi-Civita symbols. For massless free fields in
spacetime, one may write separate spinors $\phi_{AB\ldots C}$ and
$\bar{\phi}_{A'B'\ldots C'}$ for the anti-self-dual and self-dual
parts of the field respectively. These obey the general {\it massless
  free field equation}
\begin{equation}
\nabla^{AA'}\bar{\phi}_{A'\ldots C'}=0,\quad
\nabla^{AA'}\phi_{AB\ldots C}=0,
\label{masslessfreefield}
\end{equation}
where $\nabla^{AA'}$ is the spinorial translation of the covariant
derivative, and there are $2n$ indices for a spin-$n$ field. The cases
$n=0$, $1$ and $2$ correspond to solutions of a scalar
theory, gauge theory and gravity, respectively. Then
ref.~\cite{Luna:2018dpt} showed that, for vacuum solutions of Petrov
type D, the corresponding spinors were related by the {\it Weyl double
  copy} formula
\begin{equation}
  \phi_{A'B'C'D'}(x)=\frac{\phi^{(1)}_{(A'B'}(x)\phi^{(2)}_{C'D')}(x)}
  {\phi(x)}. 
  \label{WeylDC}
\end{equation}
Here $\phi^{(1,2)}_{A'B'}$ are two potentially different
electromagnetic spinors, and the brackets denote symmetrisation over
indices. Follow-up work -- including the use of the twistor methods to
be outlined below -- has established the validity of
eq.~(\ref{WeylDC}) for other Petrov
types~\cite{Godazgar:2020zbv,White:2020sfn,Chacon:2021wbr}, albeit at
linearised level only in some cases: it is only for types D and N that
Minkowski-space solutions of eq.~(\ref{masslessfreefield}) correspond
to exact solutions of the field equations.

\subsection{Twistor space and the incidence relation}
\label{sec:twistorspace}
Twistor theory provides an alternative viewpoint on solutions of
eq.~(\ref{masslessfreefield}). We start by defining twistor space
${\mathbb T}$ as the set of solutions of the {\it twistor equation}
\begin{equation}
\nabla_{A'}^{(A}\Omega^{B)}=0,
\label{twistoreq}
\end{equation}
where $\Omega^{B}$ is a spinor field. Until further notice, we will
work in {\it complexified Minkowski space} ${\mathbb M}_C$, which can
be thought of as $\mathbb{C}^{4}$ equipped with the metric
$\eta_{ab}=\text{diag}(1,-1,-1,-1)$, such that the line element in
(complex) Cartesian coordinates takes the form
\begin{equation}
d s^{2}=\eta_{ab}d x^{a}d x^{b}=(d x^{0})^{2}-(d x^{1})^{2}-(d x^{2})^{2}-(d x^{3})^{2}, \quad x^{i}\in\mathbb{C}.
\end{equation}
Given a vector $x^{a}=(x^{0},x^{1},x^{2},x^{3})$, its spinorial
representation following from eq.~(\ref{spinorial}) is
\begin{equation}
x^{AA'}=\frac{1}{\sqrt{2}}\left(\begin{matrix}
x^{0}+x^{3} & x^{1}-ix^{2} \\ 
x^{1}+ix^{2} & x^{0}-x^{3}
\end{matrix}\right),
\label{xAA'}
\end{equation}
where we have defined the Infeld-van-der-Waerden symbols as in
refs.~\cite{Penrose:1987uia,Penrose:1986ca}. We may then write the
general solution to the twistor equation of eq.~(\ref{twistoreq}) as
\begin{equation}
\Omega^A=\omega^A-ix^{AA'}\pi_{A'},
\label{twistorsol}
\end{equation}
where $\omega^A$, $\pi_{A'}$ are constant (in spacetime) spinors, that
we may combine to make a four-component ${\it twistor}$
\begin{equation}
Z^\alpha=(\omega^A,\pi_{A'}). 
\label{Zalpha}
\end{equation}
The ``location'' of a twistor in spacetime is defined to be such that
the field $\Omega^{A}$ in eq.~(\ref{twistorsol}) vanishes, which sets
up a non-local map between spacetime and twistor space known as the
{\it incidence relation}
\begin{equation}
  \omega^A=ix^{AA'}\pi_{A'}.
  \label{incidence}
\end{equation}
Given the invariance of this relation under rescalings
$Z^\alpha\rightarrow \lambda Z^\alpha$, $\lambda\in{\mathbb
  C}\setminus \{0\}$, twistors satisfying the incidence relation
correspond to points in {\it projective twistor space}
$\mathbb{PT}$. Points in spacetime correspond to complex lines
(Riemann spheres) in projective twistor space. For a point $x$ in
complexified Minkowski space, we denote the corresponding Riemann
sphere by $X\cong\mathbb{CP}^{1}$. As the components of $\pi_{A'}$
vary for a given $x^{AA'}$, they trace out all points on the Riemann
sphere $X$, so that a given point on $X$ is completely specified by a
given $\pi_{A'}$. To specify all points on $X$, we must cover it with
at least two coordinate patches, which we will label by $U_0$ and
$U_1$ in what follows.

\subsection{The Penrose transform and \u{C}ech cohomology}
\label{sec:penrose_cech}

\begin{figure}
  \begin{center}
    \scalebox{0.7}{\includegraphics{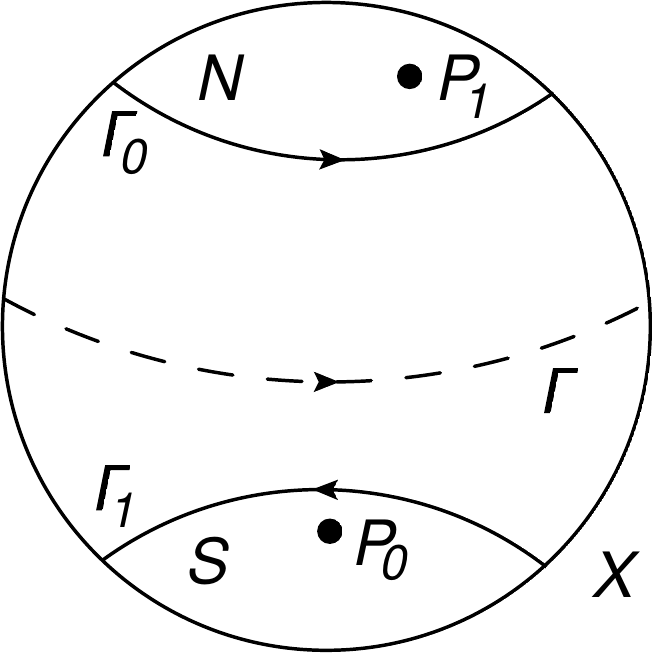}}
    \caption{The Riemann sphere $X$ in twistor space corresponding to
      a spacetime point $x$. We consider a twistor function
      $\check{f}(Z^\alpha)$ with poles $P_{i}$ contained in the regions $N$
      and $S$ around the north and south poles. The contour $\Gamma$
      separates these two poles.}
    \label{fig:sphere}
  \end{center}
\end{figure}

A key result of twistor theory is the {\it Penrose transform}, that
relates massless free fields obeying eq.~(\ref{masslessfreefield}) to
cohomological data in twistor space. More precisely, the original
formulation of the Penrose transform~\cite{Penrose:1972ia} expresses
spacetime fields as contour integrals in projective twistor space:
\begin{equation}
  \bar{\phi}_{A'B'\ldots C'}(x)=\frac{1}{2\pi i}
  \oint_\Gamma \langle\pi d\pi\rangle\pi_{A'}\pi_{B'}\ldots \pi_{C'}\check{f}(Z)\vert_{X},
  \label{Penrose}
\end{equation}
where the measure contains the inner product
\begin{equation}
\langle\pi d\pi\rangle=\pi^{A'} d\pi_{A'}=\epsilon^{A'B'}\pi_{B'} d\pi_{A'},
\label{innerprod}
\end{equation}
and the notation $\vert_{X}$ denotes restriction to the twistor line
$X\simeq\mathbb{CP}^{1}$ associated with $x$ by the incidence relation
(\ref{incidence}). Furthermore, the contour $\Gamma$ is defined on
$X$, such that it separates any poles of $\check{f}(Z^\alpha)$. All of
the functions considered in this paper will have at most two poles,
and we may parametrise the sphere such that they are contained in two
regions $N$ and $S$ around the north and south poles, as shown in
figure~\ref{fig:sphere}. Let us then define the coordinate patches
\begin{equation}
U_0=X\setminus N,\quad U_1=X\setminus S. 
  \label{U01def}
\end{equation}
That is, $U_0$ ($U_1$) consists of the sphere $X$ excluding the region
$N$ ($S$), and thus contains the pole $P_0$ ($P_1$). 
By construction, the twistor function $\check{f}(Z^\alpha)$ is
holomorphic on the intersection $U_{0}\cap U_{1}$. Also, for
eq.~(\ref{Penrose}) to make sense as an integral in {\it projective}
twistor space, the integrand plus measure must be invariant under
rescalings of $Z^\alpha$. Thus, $\check{f}(Z^\alpha)$ must be homogeneous of
degree $-2n-2$. However, one is clearly free to redefine the function
$\check{f}(Z^\alpha)$ up to contributions that vanish upon performing the
contour integral. That is, if $h_{0}$ ($h_{1}$) is a holomorphic
function on $U_{0}$ ($U_{1}$), then the Penrose transform
(\ref{Penrose}) is invariant under
\begin{equation}
 \check{f}\rightarrow\check{f}+h_{0}-h_{1}.
 \label{RepreCech}
\end{equation}
In simple terms, this corresponds to adding additional functions with
poles on only one side of the contour $\Gamma$, such that one may
always choose to close the contour in a region with no poles. The
freedom of eq.~(\ref{RepreCech}) means that it is not correct to
regard $\check{f}(Z^\alpha)$ as a function, but as a representative of
a {\it \u{C}ech cohomology class}. We refer the reader to
refs.~\cite{Penrose:1986ca,Huggett:1986fs} for excellent pedagogical
reviews of \u{C}ech cohomology in the present context, with a brief
summary as follows. Given an open cover $\{U_i\}$ of some space $X$,
one may consider a {\it $p$-cochain} $f_{i_0 i_1\ldots i_p}$,
consisting of (for our purposes) a function living on the intersection
$U_{i_0}\cap U_{i_1}\ldots \cap U_{i_p}$, where an ordering of the
intersection of the sets is implied, such that $f_{i_0 i_1\ldots i_p}$
is defined to be antisymmetric in all indices. Note that $f_i$ (a
$0$-cochain) is simply a function defined in the single patch
$U_i$. We may further define the {\it coboundary operator} $\delta_p$,
that acts on the set of $p$-cochains to make $(p+1)$-cochains:
\begin{equation}
  \delta_p (\{f_{i_0\ldots i_p}\})
  =\{(p+1) \rho_{[i_{p+1}}f_{i_0\ldots i_p]}\},
  \label{deltapdef}
\end{equation}
where square brackets denote antisymmetrisation over indices, and
$\rho_i$ denotes the restriction of a quantity to the patch $U_i$. The
$(p+1)$-cochains generated in this manner are referred to as {\it
  coboundaries}. Furthermore, cochains satisfying $\delta_p
f_{i_0\ldots i_p}=0$ are called {\it cocycles}, and we can reinterpret
the transformation of eq.~(\ref{RepreCech}) in this language. First,
note that all the quantities that appear are holomorphic functions on
the intersection $U_0\cap U_1$. Restricting to this intersection, we
may write the first term on the right-hand side, with \u{C}ech indices
made explicit, as $\check{f}_{01}$. Given that there are no triple
intersections for our cover, one automatically has $\delta_p
\check{f}_{01}=0$, so that $\check{f}_{01}$ is in fact a
cocycle\footnote{For more general covers, consistency of closed
contour integrals on triple intersections implies that the cocycle
condition $\delta_p f_{ij}=0$ indeed holds for all such quantitites
appearing in eq.~(\ref{Penrose}). See e.g. ref.~\cite{Penrose:1986ca}
for a discussion of this point.}. The function $h_0$ (defined on the
intersection) stems from a function that is holomorphic throughout the
whole of $U_0$, such that it has the form
\begin{equation}
  h_0=\rho_1 H_0,
  \label{h0H0}
\end{equation}
where $H_0$ is a 0-cochain on $U_0$. Similar reasoning applies to
$h_1$, such that we may rewrite eq.~(\ref{RepreCech}) with formal
\u{C}ech indices made explicit:
\begin{equation}
  \check{f}_{01}\rightarrow \check{f}_{01}+\rho_1 H_0-\rho_0 H_1.
\label{RepreCech2}
\end{equation}
Comparison of the latter two terms with eq.~(\ref{deltapdef}) shows
that this transformation consists of modifying the $1$-cocycle 
$\check{f}_{01}$ by a coboundary.\\

Cocycles and coboundaries both form groups, denoted by
$Z^p(\{U_i\},{\cal S})$ and $B^p(\{U_i\},{\cal S})$, where ${\cal S}$
denotes the type of function\footnote{More precisely, ${\cal S}$
denotes the so-called sheaf to which the functions belong. The
\u{C}ech cohomology described here is an approximation to sheaf
cohomology, that is sufficient for our purposes.}. We will be
concerned with holomorphic functions of homogeneity $(-2n-2)$ for a
spin-$n$ spacetime field, which we denote by ${\cal S}={\cal
  O}(-2n-2)$. Then one may define the $p^{\rm th}$ {\it \u{C}ech
  cohomology group}
\begin{equation}
  \check{H}^p(\{U_i\},{\cal S})=\frac{Z^p(\{U_i\},{\cal S})}
        {B^p(\{U_i\},{\cal S})}.
        \label{Hpdef}
\end{equation}
Elements of this group are {\it \u{C}ech cohomology classes},
consisting of cocycles that are equivalent up to addition of
coboundaries. Equation~(\ref{RepreCech2}) tells us that the quantity
appearing in the Penrose transform of eq.~(\ref{Penrose}) is indeed a
representative of a \u{C}ech cohomology class, and thus an element of
the group\footnote{Our notation here reminds us that we are
  considering cohomology groups defined on projective twistor space,
  but suggests that they will be independent of the particular cover
  $\{U_i\}$ used. That this is indeed the case follows from the fact
  that the cover $\{U_i\}$ used throughout the paper is a so-called
  {\it Leray cover}, such that \u{C}ech cohomology groups are
  isomorphic to the relevant sheaf cohomology groups.}
$\check{H}^{1}(\mathbb{PT}, \mathcal{O}(-2n-2))$.

\subsection{The Penrose transform and Dolbeault cohomology}
\label{sec:penrose_dolbeault}

An alternative formulation for the Penrose transform exists, which
uses the language of differential forms~\cite{WoodhouseTN} (see
refs.~\cite{Woodhouse:1985id,Adamo:2017qyl} for pedagogical
reviews). In general on a complex manifold ${\cal M}$ with complex
coordinates $z^i$, one may decompose differential forms into
(anti-)holomorphic parts, such that $\Omega^{p,q}({\cal M})$ denotes
the space of so-called $(p,q)$ forms
\begin{equation}
  \omega=\omega_{a_1\ldots a_p\bar{a}_1\ldots \bar{a}_q }
  dz^{a_1}\wedge \ldots \wedge dz^{a_p}\wedge d\bar{z}^{\bar{a}_1}\wedge
  \ldots \wedge d\bar{z}^{\bar{a}_q},
  \label{omegadef}
\end{equation}
where the bar on coordinates denotes complex conjugation. The exterior
derivative operator ${\rm d}$ can then be split as follows:
\begin{equation}
  {\rm d}=\partial+\bar{\partial},
  \label{dsplit}
\end{equation}
where the {\it Dolbeault operators} $\partial$ and $\bar{\partial}$
act on a $(p,q)$ form to give a $(p+1,q)$ and $(p,q+1)$ form
respectively, and are individually nilpotent
($\partial^2=\bar{\partial}^2=0$). Projective twistor space is a
complex manifold, where the precise definition of complex conjugation
depends on the signature of the spacetime we are working in. That is,
different real slices of complexified Minkowski spacetime lead to
different types of conjugation in twistor space. However, once a given
choice has been made, we may introduce the Dolbeault operator
\begin{equation}
\bar{\partial}=d \bar{Z}^{\alpha}\frac{\partial}{\partial\bar{Z}^{\alpha}},
\label{ComplexStructure}
\end{equation}
and use it to define holomorphic quantities $h$ by
$\bar{\partial}h=0$. Then the Penrose transform may be written
as~\cite{WoodhouseTN}
\begin{equation}
\phi_{A'B'\ldots C'}(x)=\frac{1}{2\pi i}\int_{X}\langle\pi d\pi\rangle\wedge\pi_{A'}\pi_{B'}\ldots \pi_{C'}f(Z)\vert_{X},
\label{Penrose2}
\end{equation}
which has a number of differences in comparison with
eq.~(\ref{Penrose}). The quantity $d\pi_{A'}$ appearing in the measure
is now to be regarded as a (1,0) form, and the integration is over the
whole Riemann sphere $X$, rather than over a contour. For this
integral to make sense, the holomorphic quantity $f(Z)$ must be a
(0,1) form which, given we are in projective twistor space as before,
must again have homogeneity $-2n-2$ for a spin-$n$ field:
\begin{gather}
f\in \Omega^{0,1}(\mathbb{PT},\mathcal{O}(-2n-2)),\qquad 
f(\lambda Z)=\lambda^{-2n-2}f(Z),\qquad \bar{\partial}f=0.
\end{gather}
Similarly to eq.~(\ref{Penrose}), there is a redundancy in how one
chooses $f(Z)$: if we redefine it according to
\begin{equation}
f\rightarrow f+\bar{\partial}g,
\label{RepresTwistor}
\end{equation}
for some $g\in\Omega^{0}(\mathbb{PT},\mathcal{O}(-2s-2))$, and where
$\bar{\partial}g$ is globally defined on $X$, the second term will
vanish as a total derivative on the Riemann sphere when inserted in
eq.~(\ref{Penrose2}). Furthermore, the additional term preserves the
holomorphic property $\bar{\partial}f(Z)=0$, by nilpotency of
$\bar{\partial}$. In general, the set of $(p,q)$ forms on a manifold
${\cal M}$ satisfying $\bar{\partial}h=0$ are called
$\bar{\partial}$-closed, and form a group under addition denoted by
$Z^{p,q}_{\bar{\partial}}({\cal M})$. Forms of the form
$h=\bar{\partial} g$ are called $\bar{\partial}$-exact, and form the
group $B^{p,q}_{\bar{\partial}}({\cal M})$. One may then define the
$(r,s)^{\rm th}$ {\it Dolbeault cohomology group}
\begin{equation}
  H_{\bar{\partial}}^{p,q}({\cal M})
  =\frac{Z_{\bar{\partial}}^{p,q}({\cal M})}
  {B_{\bar{\partial}}^{r,s}({\cal M})}.
  \label{HDolbeault}
\end{equation}
Elements of this group are {\it Dolbeault cohomology classes}, namely
$\bar{\partial}$-closed $(p,q)$ forms that are defined only up to
arbitrary additions of $\bar{\partial}$-exact forms. It follows from
these definitions that the twistor (0,1)-form $f(Z)$ appearing in
eq.~(\ref{Penrose2}) is a representative element of the Dolbeault
cohomology group
\begin{equation}
H^{0,1}_{\bar{\partial}}(\mathbb{PT},\mathcal{O}(-2n-2)),
\end{equation}
where our enhanced notation relative to eq.~(\ref{HDolbeault}) makes
clear that we are considering holomorphic $(0,1)$ forms of a certain
homogeneity only.

\subsection{Connection between Dolbeault and \u{C}ech descriptions}
\label{sec:connection}

The previous sections provide two different descriptions of the
cohomological identification of spacetime fields implied by the
Penrose transform. Let us now examine the relationship between them,
where we will follow the arguments presented in
e.g. refs.~\cite{Ward:1990vs,Jiang:2008xw}. We will consider
explicitly the situation of figure~\ref{fig:sphere} in the Dolbeault
approach, so that $f(Z)|_X$ is a (0,1) form that is holomorphic
everywhere apart from singularities at $P_0$ and $P_1$. Then the
Dolbeault representative $f(Z)$ associated with a given \u{C}ech
representative $\check{f}(Z)$ may be defined as follows. First, given
our cover $(U_0,U_1)$ of $X$, we may choose a partition of unity
$\{\eta_i\}$, where each $\eta_i$ is defined in $U_i$, subject to
\begin{equation}
  \sum_i \eta_i=1.
  \label{partition}
\end{equation}
We may thus write
\begin{equation}
  \eta_0=\eta,\quad \eta_1=1-\eta,
\end{equation}
and also define
\begin{equation}
f_i=\sum_j \check{f}_{ij}\eta_j
\label{fidef}
\end{equation}
in $U_i$, so that we have explicitly
\begin{equation}
  f_0=(1-\eta)\check{f}_{01},\quad f_1=\eta
  \check{f}_{10}=-\eta \check{f}_{01}.
\label{f01def}
\end{equation}
Then the desired Dolbeault representatives are given by
\begin{equation}
  f(Z)=\{\bar{\partial}f_i \}.
  \label{fZdef}
\end{equation}
This satisfies $\bar{\partial}f(Z)=0$ by construction. Furthermore, on
the intersection $U_0\cap U_1$, one has (via eq.~(\ref{f01def})) 
\begin{equation}
  \bar{\partial}f_0-\bar{\partial}f_1=\bar{\partial}\check{f}_{01}=0,
  \label{global}
\end{equation}
so that $f(Z)$ is indeed uniquely defined globally on $X$. To check
these identifications, it is instructive to see how the Penrose
transform of eq.~(\ref{Penrose2}) reduces to that of
eq.~(\ref{Penrose}). First, note that one may write the integral over
the Riemann sphere $X=U_0\cup U_1$ in eq.~(\ref{Penrose2}) as
\begin{displaymath}
  \left[\int_{U_0}+\int_{U_1}-\int_{U_0\cap U_1}\right]
  \langle\pi d\pi\rangle\wedge\pi_{A'}\pi_{B'}\ldots \pi_{C'}f(Z)\vert_{X}.
\end{displaymath}
In the third term, the integrand will contain $\bar{\partial}\check{f}_{01}$
evaluated on the intersection (i.e. away from the poles at $P_0$ and
$P_1$), which is zero. Thus, we need only consider the first two
terms. Substituting the results of eq.~(\ref{f01def}), we may rewrite
them using Stokes' theorem to give 
\begin{align}
 \left( (1-\eta)\oint_{\Gamma_0} \langle \pi d\pi\rangle \pi_{A'}\pi_{B'}\ldots
 \pi_{C'} \check{f}_{01}(Z)\right)\notag\\
 -\left(\eta \oint_{\Gamma_1} \langle \pi d\pi \rangle
 \pi_{A'}\pi_{B'}\ldots \pi_{C'}\check{f}_{01}(Z)\right),
 \label{Stokes}
\end{align}
where $\Gamma_i$ is the oriented boundary of $U_i$. These boundaries
are depicted in figure~\ref{fig:sphere}, and the absence of poles
between each $\Gamma_i$ and $\Gamma$ imply that one may write
\begin{displaymath}
  \oint_{\Gamma_0}\equiv\oint_{\Gamma},\quad
  \oint_{\Gamma_1}\equiv-\oint_{\Gamma},
\end{displaymath}
where one must take the opposite orientation of $\Gamma_1$ on the
sphere into account. The remaining integral over $d\pi$ can be
interpreted as a conventional contour integral, such that
eq.~(\ref{Stokes}) reproduces the Penrose transform of
eq.~(\ref{Penrose}), as required. We have here shown how to go from
the \u{C}ech representative of a twistor function to the corresponding
Dolbeault representative. For a discussion of how to go the other way,
we refer the reader to e.g. refs.~\cite{Jiang:2008xw,Ward:1990vs}.\\

As well as showing how Dolbeault representatives may be defined from
their \u{C}ech counterparts, we may also reinterpret the cohomological
freedom. It follows from eq.~(\ref{fidef}) that redefining a \u{Cech}
representative by
\begin{displaymath}
  \check{f}_{ij}\rightarrow \check{f}_{ij}+h_i-h_j
\end{displaymath}
amounts to redefining the Dolbeault representative according to
\begin{equation}
  f(Z)\rightarrow f(Z)-\bar{\partial}\left(\sum_i h_i\eta_i\right).
\label{fijshift}
\end{equation}
Above, we have used an arbitrary partition of unity on our cover
$(U_0,U_1)$. We can simplify things, however, by choosing a trivial
partition in which $\eta=0$. Then the Dolbeault Penrose transform can
be carried out by integrating solely over the patch $U_0$, even though
this does not cover the entire sphere.

\subsection{The twistor double copy}
\label{sec:twistcopy}

Having reviewed various aspects of the Penrose transform, let us now
turn our attention to the Weyl double copy of eq.~(\ref{WeylDC}),
connecting scalar, gauge and gravity fields in spacetime. As was
recently presented in refs.~\cite{White:2020sfn,Chacon:2021wbr}, it is
possible to derive this relationship from the Penrose transform of
eq.~(\ref{Penrose}) (i.e. in the \u{C}ech language). The procedure
involves choosing holomorphic twistor quantities $\check{f}(Z)$,
$\check{f}^{(i)}_{\rm EM}(Z)$ of homogeneity $-2$ and $-4$ respectively,
such that they correspond to scalar and EM fields in spacetime
respectively. One may then form the product
\begin{equation}
  \check{f}_{\rm grav.}(Z)=\frac{\check{f}_{\rm EM}^{(1)}(Z)
    \check{f}_{\rm EM}^{(2)}(Z)}
  {\check{f}(Z)},
  \label{twistorcopy}
\end{equation}
which has homogeneity $-6$ by construction. This corresponds to a
gravity field in spacetime, and
refs.~\cite{White:2020sfn,Chacon:2021wbr} presented choices for the
various functions appearing on the right-hand side of
eq.~(\ref{twistorcopy}) such that the spacetime fields obtained from
eq.~(\ref{Penrose}) obey the Weyl double copy of
eq.~(\ref{WeylDC}). For the original type D Weyl double copy of
ref.~\cite{Luna:2018dpt}, it is sufficient to choose functions of the
form
\begin{equation}
  \check{f}_m(Z)=\left[Q_{\alpha\beta}Z^\alpha Z^\beta\right]^{-m},
  \label{fmdef}
\end{equation}
for some constant dual twistor $Q_{\alpha\beta}$, and where $m=1$ and
$2$ for the scalar and EM cases respectively. This is a quadratic form
in twistor space, and implies the presence of two poles on the Riemann
sphere $X$ corresponding to a given spacetime point $x$. The scalar,
gauge and gravity fields linked by eq.~(\ref{twistorcopy}) then share
the same poles. These poles give rise to the principal spinors of
their respective spacetime fields, so that one obtains a geometric
interpretation of how kinematic information is inherited between
different theories in spacetime. Furthermore,
ref.~\cite{Chacon:2021wbr} provided examples of non-type D solutions
(albeit at linearised level due to the limitations of the Penrose
transform), showing that at the very least the twistor double copy
provides a highly convenient book-keeping device for constructing
spacetime examples of the Weyl double copy.\\

However, there is an obvious deficiency of eq.~(\ref{twistorcopy}),
discussed in detail in ref.~\cite{Chacon:2021wbr}. As reviewed above,
the quantities appearing in eq.~(\ref{twistorcopy}) are not in fact
functions, but representatives of cohomology classes, which may in
principle be subjected to the equivalence of transformations of
eq.~(\ref{RepreCech}). The product of eq.~(\ref{twistorcopy}), in
being a non-linear relationship, clearly violates this
invariance. Upon redefining the scalar and gauge theory quantities
$\check{f}(Z)$ and $\check{f}^{(i)}_{\rm EM}$ {\it before} forming the
product, one would obtain a different gravity solution in
general. This does not matter from the point of view of deriving the
Weyl double copy: all that is required is that we find suitable
quantitites in twistor space that correspond to the desired spacetime
relationship. However, if the classical double copy is to be given a
genuinely twistorial interpretation, we need a prescription for
picking a ``special'' representative for each cohomology class, that
eliminates any ambiguity in the double copy procedure. This has been
discussed recently in ref.~\cite{Adamo:2021dfg}, which focused on
purely radiative spacetimes, namely those that can be completely
determined by data at future null infinity. It is known that this
characteristic data can be used, in either gauge theory or gravity, to
uniquely fix a Dolbeault representative in the Penrose
transform~\cite{MasonTN}. Thus, for such spacetimes a natural
mechanism arises for fixing the ambiguities inherent in
eq.~(\ref{twistorcopy}). It is not immediately clear, however, how to
generalise this argument to non-radiative spacetimes, and thus we will
present alternative arguments in what follows.

\section{The twistor double copy in the Dolbeault approach}
\label{sec:Dolbeault}

In the previous section, we reviewed the twistor double copy of
eq.~(\ref{twistorcopy}), based on the Penrose transform of
eq.~(\ref{Penrose}), in which all twistor functions are to be
interpreted as representatives of \u{C}ech cohomology classes. Let us
now see how one can instead formulate the same idea within the
framework of Dolbeault cohomology. We will begin by studying a
particularly simple example of solutions of
eq.~(\ref{masslessfreefield}).

\subsection{Momentum eigenstates}
\label{sec:planewaves}

Momentum eigenstates in spacetime are characterised by a given null
momentum with spinorial translation $p_a\rightarrow \tilde{p}_A
p_{A'}$, and are a special case of plane waves. The solution of
eq.~(\ref{masslessfreefield}) corresponding to such a wave can then be
written as
\begin{equation}
  \bar{\phi}_{A'B'\ldots C'}(x)
  =p_{A'}p_{B'}\ldots p_{C'}
  e^{ip\cdot x},
   \label{planewaves}
\end{equation}
where the basic phase factor $e^{ip\cdot
  x}=e^{i\tilde{p}_Ap_{A'}x^{AA'}}$ is dressed by an appropriate
number of spinors $p_{A'}$, according to the spin of the relevant
field. We must then be able to find a (0,1)-form $f(Z)$ in projective
twistor space that, when restricted to the Riemann sphere $X$ and
substituted into eq.~(\ref{Penrose2}), yields the spacetime field of
eq.~(\ref{planewaves}) for a given spin. This (0,1) form will be
defined only up to the addition of an arbitrary
$\bar{\partial}$-closed (0,1) form, and a suitable Dolbeault
representative for a plane wave of helicity $h$ can be written as (see
e.g. ref.~\cite{Adamo:2017qyl})
\begin{equation}
  f^{[h]}=\left(\frac{\langle a p\rangle}{\langle a\pi\rangle}
  \right)^{-2h+1}\bar{\delta}(\langle\pi p\rangle)
  \exp\left[\frac{\langle a p\rangle}{\langle a\pi\rangle}
    [\omega \tilde{p}]\right],
  \label{planewaverep}
\end{equation}
where $a_{A'}$ is an arbitrary constant Weyl spinor, and we have
introduced a holomorphic delta function to be inserted into our
Penrose transform integral, which may be further decomposed using the
useful identity\footnote{Our convention for the holomorphic delta
function differs from that of ref.~\cite{Adamo:2017qyl} in that we have not included a factor of $(2\pi i)^{-1}$. The reason is that this has already been included in our Penrose transform definition of eq.~(\ref{Penrose2}). }
\begin{equation}
  \bar{\delta}(u)=\bar{\partial}
  \left(\frac{1}{u}\right).
  \label{deltabar}
\end{equation}
We have also introduced the inner product
\begin{equation}
[\omega\tilde{p}]=\omega^A \tilde{p}_A,
\end{equation}
where $\omega^A$ is the Weyl spinor appearing in $Z^\alpha$ according
to eq.~(\ref{Zalpha}). To see that eq.~(\ref{planewaverep}) indeed
reproduces eq.~(\ref{planewaves}), regardless of the choice of
$a_{A'}$, we may parametrise $X$ in eq.~(\ref{Penrose2}) by choosing 
\begin{equation}
  \pi_{A'}=b_{A'}+z a_{A'}\quad\Rightarrow\quad \langle \pi d\pi\rangle=
  -\langle ab\rangle dz,\quad \pi_{A'}=\frac{\langle ab\rangle}
{\langle ap\rangle}p_{A'},
  \label{piparam}
\end{equation}
where $b_{A'}$ is another constant spinor such that $\langle
ab\rangle\neq 0$, and we have used the delta function condition in the
third equation. Without loss of generality, let us choose this
parametrisation to correspond to the patch $U_0$ discussed in the
previous section, such that this contains the support of the
holomorphic delta function. Equation~(\ref{planewaverep}) then
becomes, after restriction to the Riemann sphere $X$,
\begin{equation}
  f^{[h]}\Big|_{X}=\frac{\langle ap \rangle^{-1}}{2\pi i}
\left(\frac{\langle a p\rangle}{\langle a b\rangle}
  \right)^{-2h+1}\bar{\partial}\left(\frac{1}
{z+\frac{\langle bp\rangle}{\langle ap \rangle}}\right)
  \exp\left[ix^{AA'}p_{A'}\tilde{p}_A\right],
  \label{planewaverep2}
\end{equation}
where the incidence relation of eq.~(\ref{incidence}) has been
used. Substituting this into eq.~(\ref{Penrose2}), one finds
\begin{displaymath}
p_{A'}\ldots p_{C'}\exp\left[ix^{AA'}p_{A'}\tilde{p}_A\right]
\frac{1}{2\pi i}\int_{U_0} dz\wedge \bar{\partial}\left(
\frac{1}{z+\frac{\langle bp\rangle}{\langle ap\rangle}}\right),
\end{displaymath}
so that carrying out the integral using Stoke's theorem yields
eq.~(\ref{planewaves}) as required. Note that we integrated only over
$U_0$ above, rather than the complete Riemann sphere $X$. In order to
complete the latter, as per the discussion in
section~\ref{sec:connection}, one should also integrate over a second
patch $U_1$. However, by construction this has be taken so as not to
contain the support of the holomorphic delta function, and thus the
further integration will not affect the above result, as expected
given that we have already recovered the plane wave spacetime field of
eq.~(\ref{planewaves}).\\

Now let us examine plane waves of different helicity, and note that we
may choose to rewrite eq.~(\ref{planewaverep2}) (before restriction to
$X$) as
\begin{equation}
f^{[h]}=\bar{\partial} F^{[h]},
\label{Fhdef1}
\end{equation}
where
\begin{equation}
F^{[h]}=\langle ap\rangle^{-1}\left(
\frac{\langle ap \rangle}{\langle ab \rangle}\right)^{-2h+1}
\frac{1}{z+\frac{\langle bp\rangle}{\langle ap\rangle}}
\exp\left[i\frac{\langle ap\rangle}{\langle ab\rangle}[\omega
\tilde{p}]\right].
\label{Fhdef2}
\end{equation}
It is then straightforward to verify the relationship
\begin{equation}
\label{DC_from_partia_PWl}
F^{[h+h']}=\frac{F^{[h]}F^{[h']}}{F^{[0]}},
\end{equation}
which can be interpreted as follows. Choosing $h=h'=1$, one finds that
eq.~(\ref{Fhdef1}) applied to $F^{[0]}$ and $F^{[1]}$ yields Dolbeault
representatives associated with scalar and gauge theory respectively,
such that the Penrose transform of eq.~(\ref{Penrose2}) gives
spacetime scalar and photon plane
waves. Equation~(\ref{DC_from_partia_PWl}), after substitution into
eq.~(\ref{Fhdef1}), yields a Dolbeault representative for a gravity
wave. From eq.~(\ref{planewaves}), the resulting spacetime fields are
then precisely related by the Weyl double copy of
eq.~(\ref{WeylDC}). Thus, eq.~(\ref{DC_from_partia_PWl}) is a
twistor-space expression of the Weyl double copy, that can be used to
generate the Dolbeault representative for a gravity solution, from
similar representatives in scalar and gauge theory. Of course, plane
waves are very special solutions, and it is perhaps not clear that the
procedure of eq.~(\ref{DC_from_partia_PWl}) generalises to a wider
class of solutions. That it indeed generalises in fact follows from
the ideas reviewed in section~\ref{sec:connection}, as we now discuss.

\subsection{The twistor double copy from Dolbeault representatives}
\label{sec:dolbeaultDC}

The previous section suggests the following general
prescription. Consider Dolbeault representatives
\begin{equation}
f(Z)=\bar{\partial}F(z),\quad f^{(l)}_{\rm EM}(Z)=\bar{\partial}
F^{(l)}_{\rm EM}
\label{fF}
\end{equation}
defined locally on some patch $U_i$, corresponding to scalar and
electromagnetic fields respectively. Then one can form a gravitational
Dolbeault representative on $U_i$ by
\begin{equation}
f_{\rm grav.}=\bar{\partial}\left[
\frac{F^{(1)}_{\rm EM}(Z)F^{(2)}_{\rm EM}(Z)}{F(Z)}
\right].
\label{Fgrav}
\end{equation}
Our claim is then that suitable representatives may be chosen so that
the corresponding spacetime fields obtained from eq.~(\ref{Penrose2})
are related by the Weyl double copy of eq.~(\ref{WeylDC}). To see why,
note that one may choose \u{C}ech representatives in
eq.~(\ref{twistorcopy}) so as to obtain a gravitational \u{C}ech
representative, where the corresponding fields obey the Weyl double
copy. We may then convert each \u{C}ech representative to a Dolbeault
representative using eqs.~(\ref{fidef}, \ref{fZdef}). To simplify this
procedure, we may choose a trivial partition of unity, such that
$\eta=0$. We may then carry out the Penrose transform of
eq.~(\ref{Penrose2}) by integrating only over the patch $U_0$, and
such that the functions appearing in eq.~(\ref{fF}) are simply given
by
\begin{equation}
F(z)=\check{f}(Z),\quad F^{(l)}_{\rm EM}=\check{f}^{(l)}_{\rm EM}.
\label{fF2}
\end{equation}
Thus, the Dolbeault double copy formula of eq.~(\ref{Fgrav}) is
ultimately a rewrite of the \u{C}ech formula, where the latter is
converted to a (0,1) form by the action of the Dolbeault operator
$\bar{\partial}$. Upon integrating the Penrose transform over $U_0$, a
non-zero result survives provided the quantity in the square brackets
in eq.~(\ref{Fgrav}) has a pole in the patch $U_0$. To clarify our
rather abstract discussion, we now present some illustrative examples.

\subsubsection{Elementary states}

In the \u{C}ech language, {\it elementary states} are holomorphic
twistor functions consist of simple ratios of factors such as
$(A_\alpha Z^\alpha)$, where $A_\alpha$ is a constant dual
twistor. They were originally studied as potential twistor-space
wavefunctions for scattering particles (see
e.g. refs.~\cite{Penrose:1972ia,Penrose:1986ca}), and have since been
reinterpreted as giving rise to novel knotted solutions of gauge and
gravity
theory~\cite{Dalhuisen:2012zz,Swearngin:2013sks,deKlerk:2017qvq,Thompson:2014owa,Thompson:2014pta,Sabharwal:2019ngs}. In
refs.~\cite{White:2020sfn,Chacon:2021wbr}, elementary states were used
to construct examples of Weyl double copies where the gravity solution
had arbitrary Petrov type, albeit at linearised level. Given such an
elementary state, we may form a suitable Dolbeault representative as
in eqs.~(\ref{fF}, \ref{fF2}), and thus consider the following family
of (0,1) forms:
\begin{equation}
f^{(a,b)}(Z)=\bar{\partial}\left(\frac{1}{(A_\alpha Z^\alpha)^{a+1}
(B_{\beta}Z^{\beta})^{b+1}}
\right),
\label{fabdef}
\end{equation}
where $2n=a+b$ for a spin-$n$ field, and the dual twistors
$A_{\beta}=(A_{A},A^{A'}),B_{\alpha}=(B_{B},B^{B'})$. Upon restricting
to the Riemann sphere $X$ of a given spacetime point $x$, let us
choose a cover $(U_0,U_1)$ such that the pole in $A\cdot Z$ ($B\cdot
Z$) lies in the patch $U_0$ but not $U_1$ ($U_1$ but not $U_0$). In
$U_0$, eq.~(\ref{fabdef}) may then be written as
\begin{equation}
f^{(a,b)}(Z)\Big|_{U_0}=\frac{1}{(B_{\beta}Z^{\beta})^{b+1}}
\bar{\partial}\left(\frac{1}{(A_\alpha Z^\alpha)^{a+1}
}
\right).
\label{fabdef2}
\end{equation}
The product between the twistor $Z^{\beta}$ and the dual twistor
$A_{\beta}$ is given by
\begin{eqnarray}
A_{\beta}Z^{\beta}\vert_{X}=ix^{AA'}A_{A}\pi_{A'}+A^{A'}\pi_{A'}
=(ix^{AA'}A_{A}+A^{A'})\pi_{A'}\equiv\langle\mathcal{A}\pi\rangle,
\label{RobinsonA}
\end{eqnarray}
where we have introduced the {\it Robinson field}
\begin{equation}
\mathcal{A}^{A'}=ix^{AA'}A_{A}+A^{A'},
\label{RobinsonA2}
\end{equation}
and used eq.~(\ref{incidence}). Similarly, one may write 
\begin{equation}
B_{\alpha}Z^{\alpha}\vert_{X}\equiv\langle\mathcal{B}\pi\rangle,\quad
\mathcal{B}^{A'}=ix^{AA'}B_{A}+B^{A'},
\label{RobinsonB}
\end{equation}
such that eq.~(\ref{fabdef2}) becomes
\begin{equation}
f(Z)\Big|_{U_0}
=\frac{1}{\langle\mathcal{B}\pi\rangle^{b+1}}\bar{\partial}\left(\frac{1}{\langle\mathcal{A}\pi\rangle^{a+1}}\right).
\end{equation}
Under the Penrose transform, and using our trivial partition of unity,
we have
\begin{equation}
\phi_{A'_{1}\ldots A'_{2n}}(x)=\frac{1}{2\pi i}\int_{U_0}\langle\pi d\pi\rangle\wedge\pi_{A'_{1}}\ldots \pi_{A'_{2n}}\frac{1}{\langle\mathcal{B}\pi\rangle^{b+1}}\bar{\partial}\left(\frac{1}{\langle\mathcal{A}\pi\rangle^{a+1}}\right),
\label{elstate1}
\end{equation}
which may be explicitly evaluated by making the following
parametrisation for $U_0$:
\begin{equation}
\pi_{A'}(z)=\mathcal{A}_{A'}+z\mathcal{B}_{A'}.
\label{CoordinateLambda}
\end{equation}
Note that our requirement that $U_0$ does not contain the pole at
$\langle \mathcal{B}\pi\rangle=0$ implies
$\langle\mathcal{B}\mathcal{A}\rangle\neq0$. Substituting
eq.~(\ref{CoordinateLambda}) into eq.~(\ref{elstate1}) then yields
\begin{align}
\phi_{A'_{1}\ldots A'_{2n}}(x)&=\frac{1}{2\pi i}
\frac{(-1)^{a}}{\langle \mathcal{B}\mathcal{A}\rangle^{a+b+1}}
\int_{U_0}
dz \,\pi_{A'_{1}}(z)\ldots 
\pi_{A'_{2n}}(z)\bar{\partial}
\left(\frac{1}{z^{a+1}}\right)\notag\\
&=\frac{1}{2\pi i}
\frac{(-1)^{a}}{\langle \mathcal{B}\mathcal{A}\rangle^{a+b+1}}
\oint_{\partial U_0}\frac{dz}{z^{a+1}} \,\pi_{A'_{1}}(z)\ldots 
\pi_{A'_{2n}}(z),
\label{PenroseTransComplex}
\end{align}
where Stokes' theorem has been used in the second line. Taking the
residue of the pole at $z=0$, one finds
\begin{align}
\phi_{A'_{1}\ldots A'_{2n}}(x)&=\frac{(-1)^{a}}{\langle\mathcal{BA}\rangle^{a+b+1}}\frac{1}{a!}\lim_{z\rightarrow 0}\frac{ d ^{a}}{ d z^{a}}
\left[\pi_{A'_{1}}(z)\ldots\pi_{A'_{2n}}(z)\right]\nonumber\\
&=\frac{(-1)^{a}}{\langle\mathcal{BA}\rangle^{a+b+1}}\binom{a+b}{a}\mathcal{A}_{(A'_{1}}\ldots\mathcal{A}_{A'_{b}}\mathcal{B}_{A'_{b+1}}\ldots\mathcal{B}_{A'_{2n})}.
\label{elstate2}
\end{align}
Special cases of this family of spacetime fields indeed obey the Weyl
double copy of eq.~(\ref{WeylDC}), as already discussed in
refs.~\cite{White:2020sfn,Chacon:2021wbr}. Indeed, this will be the
case whenever scalar, EM and gravity fields are chosen such that
eqs.~(\ref{fF}, \ref{Fgrav}) are obeyed for their Dolbeault
representatives in twistor space, as may be easily verified.

\subsubsection{General type D vacuum solutions}
\label{sec:typeD}

To go further than the previous section, we may consider the Dolbeault
representatives
\begin{equation}
f_m(Z)=\bar{\partial}F_m(Z),\quad F_m(Z)=\check{f}_m(Z),
\label{typeD}
\end{equation}
where $\check{f}_m(Z)$ consists of an inverse power of a quadratic
form, shown explicitly in eq.~(\ref{fmdef}). On the Riemann sphere
$X$, let us parametrise $U_0$ by
\begin{equation}
\pi_{A'}=(1,\xi)
\label{piparam2}
\end{equation}
i.e. such that $\pi_{0'}\neq 0$. The quadratic form will assume the
general form
\begin{equation}
Q_{\alpha\beta}Z^\alpha Z^\beta=N^{-1}(x)(\xi-\xi_0)(\xi-\xi_1)
\label{Qparam}
\end{equation}
where the normalisation factor $N^{-1}(x)$ inherits its spacetime
dependence from the incidence relation, and the pole $\xi_i$ is taken
to lie exclusively in $U_i$, as shown in figure~\ref{fig:sphere}. The
Penrose transform of eq.~(\ref{Penrose2}) for each $m$ then evaluates
to
\begin{align}
\phi_{A'\ldots D'}(x)&=-\frac{N^m(x)}{2\pi i}\int_{U_0}
d\xi \wedge \frac{(1,\xi)_{A'}\ldots (1,\xi)_{D'}}{(\xi-\xi_1)^m}
\bar{\partial}\left(\frac{1}{(\xi-\xi_0)^m}\right)\notag\\
&=-\frac{N(x)}{2\pi i}\oint_{\partial U_0}
d\xi \frac{(1,\xi)_{A'}\ldots (1,\xi)_{D'}}{(\xi-\xi_0)^m(\xi-\xi_1)^m}.
\label{Dcalc1}
\end{align}
Taking the residue of the pole at $\xi=\xi_0$, one finds spacetime
fields (for $m$=1, 2 and 3 respectively)
\begin{equation}
\phi(x)=-\frac{N(x)}{\xi_0-\xi_1},\quad
\phi_{A'B'}(x)=\frac{N^2(x)}{(\xi_0-\xi_1)^3}\alpha_{(A'}\beta_{B')},
\quad 
\phi_{A'B'C'D'}=-\frac{N^3(x)}{(\xi_0-\xi_1)^5}\alpha_{(A'}\alpha_{B'}
\beta_{C'}\beta_{D')},
\label{phifields}
\end{equation}
where we have introduced the spinors
\begin{equation}
\alpha=(1,\xi_0),\quad \beta=(1,\xi_1).
\label{alphabetadef}
\end{equation}
The fields of eq.~(\ref{phifields}) clearly obey the Weyl double copy
of eq.~(\ref{WeylDC}). Furthermore, as has been pointed in
ref.~\cite{Haslehurst}, use of a general dual twistor
$Q_{\alpha\beta}$ allows one to span the complete space of vacuum type
D solutions. We thus recover the results of
refs.~\cite{White:2020sfn,Chacon:2021wbr} in the \u{C}ech approach,
but this was in any case guaranteed by our general argument. The
examples of alternative Petrov types presented in
refs.~\cite{White:2020sfn,Chacon:2021wbr} will also generalise to the
Dolbeault approach.\\

To summarise, in this section we have introduced a general procedure
for obtaining twistor double copies in the Dolbeault formalism, which
is essentially a rewriting of the twistor double copy in the \u{C}ech
approach to twistor theory. Alas, this means that the Dolbeault
approach also suffers from the same apparent ambiguities as the
\u{C}ech double copy, which we discuss in the following section.

\subsection{Cohomology and the Dolbeault double copy}
\label{sec:problem}

As we reviewed in section~\ref{sec:review}, the twistor quantities
appearing in the Penrose transforms of eq.~(\ref{Penrose})
and~(\ref{Penrose2}) are representatives of cohomology classes. In
eq.~(\ref{Penrose}), these are \u{C}ech cohomology classes, and we saw
that the product of eq.~(\ref{twistorcopy}) is incompatible in general
with the ability to redefine each representative according to the
equivalence transformations of eq.~(\ref{RepreCech}). In the Dolbeault
language, this freedom translates to the ability to add a
$\bar{\partial}$-exact form to each representative, as expressed in
eq.~(\ref{RepresTwistor}). One may then investigate whether the
prescription of eq.~(\ref{Fgrav}) respects the ability to redefine
each Dolbeault representative according to eq.~(\ref{RepresTwistor}),
and it is straightforward to see that it does not. \\

To show this, recall that the functions appearing in eq.~(\ref{Fgrav})
are straightforwardly related to their corresponding \u{Cech}
representatives by eq.~(\ref{fF2}). Redefining the latter according to
eq.~(\ref{RepreCech}) amounts, from eq.~(\ref{fijshift}) and our
partition of unity with $\eta_0=0$, to adding $\bar{\partial}h_0(Z)$
to the corresponding Dolbeault representative. Here $h_0(Z)$ has poles
only in $U_1$. In
eq.~(\ref{Fgrav}), the conversion of the square brackets to a (0,1)
form happens {\it after} the double copy product has already taken
place. For our purposes, it is sufficient to consider equivalence
transformations of the functions appearing in the numerator of
eq.~(\ref{Fgrav}), such that one replaces eq.~(\ref{Fgrav}) with
\begin{align}
f_{\rm grav.}(Z)&\rightarrow \bar{\partial}\left[
\frac{(F^{(1)}_{\rm EM}(Z)+h_0^{(1)}(Z))(F^{(2)}_{\rm EM}(Z)+h_0^{(2)}(Z))}{F(Z)}
\right]\notag\\
&= f_{\rm grav.}(Z)+\bar{\partial}\left[\frac{h_0^{(1)}(Z)F^{(2)}_{\rm EM}(Z)+
  h_0^{(2)}(Z)F^{(1)}_{\rm EM}(Z)+h_0^{(1)}(Z)h_0^{(2)}(Z)}{F(Z)}
  \right].
  \label{Fgrav2}
\end{align}
We stress that, despite appearances, this replacement does not have
the same form as the equivalence transformation of
eq.~(\ref{RepresTwistor}): in the latter, the second term is defined
over the whole Riemann sphere, and thus vanishes as a total derivative
when integrated. By contrast, in eq.~(\ref{Fgrav2}) the second term is
defined only locally within the patch $U_0$, and thus gives a
potentially non-zero result after integration. Indeed, Stoke's theorem
implies that the contribution to the second term in eq.~(\ref{Fgrav2})
to the Penrose transform integral is 
\begin{displaymath}
  \frac{1}{2\pi i}\oint_{\partial U_0}\langle \pi d\pi\rangle
  \pi_{A'}\pi_{B'}\pi_{C'}\pi_{D'}
\left[\frac{h_0^{(1)}(Z)F^{(2)}_{\rm EM}(Z)+
  h_0^{(2)}(Z)F^{(1)}_{\rm EM}(Z)+h_0^{(1)}(Z)h_0^{(2)}(Z)}{F(Z)}
  \right].
\end{displaymath}
The terms in the square brackets have poles in $U_0$ in general, and
thus this integral will be non-zero. Thus, as in the \u{C}ech double
copy of refs.~\cite{White:2020sfn,Chacon:2021wbr}, redefining the
scalar and EM fields {\it before} forming the twistor space product
results in a different spacetime field in general. As in that case,
this is not a problem when it comes to deriving the form and scope of
the Weyl double copy, where one must simply find suitable
representatives for each field in twistor space so as to
reproduce the desired spacetime relationship. However, it would be
nice if there were a systematic way to decide which representative
should be chosen. We give one such method in the following section.

\section{Dolbeault representatives in Euclidean signature}
\label{sec:Euclidean}

As we have seen, both the \u{C}ech and Dolbeault double copies involve
forming apparently ambiguous products of twistor functions, where the
non-linear nature of this relationship is at odds with the fact that
these functions are actually representatives of cohomology classes. It
is then natural to ponder whether there are any natural ways to choose
``special'' representatives of each class, so that the procedure can
be made unambiguous. One such procedure has been presented recently in
ref.~\cite{Adamo:2021dfg}, which focused on radiative spacetimes. Here
we give a different procedure that works for all examples considered
in this paper, provided one uses Euclidean signature in
spacetime. This allows the use of known methods from complex analysis
that can indeed pick out special representatives of Dolbeault
cohomology classes. For reviews of twistor theory in Euclidean
signature, see refs.~\cite{Woodhouse:1985id,Adamo:2017qyl}, the latter
of which inspires our review of relevant material below.\\

The spinorial translation of a spacetime point $x^a$ has been given in
eq.~(\ref{xAA'}). One may impose Euclidean signature by defining the
hat-operation
\begin{equation}
\hat{x}^{AA'}=\frac{1}{\sqrt{2}}\left(\begin{matrix}
\bar{x}^{0}-\bar{x}^{3} & -\bar{x}^{1}+i\bar{x}^{2} \\ 
-\bar{x}^{1}-i\bar{x}^{2} & \bar{x}^{0}+\bar{x}^{3}
\end{matrix} \right),
\end{equation}
where the bar denotes complex conjugation. Demanding that
$x^{AA'}=\hat{x}^{AA'}$ yields the constraints $x^0\in\mathbb{R}$ and
\begin{displaymath}
  x^l=iy^l,\quad y_l\in\mathbb{R},\quad l\in\{1,2,3\},
\end{displaymath}
such that
\begin{equation}
  x_a x^a = (x^0)^2+(y^1)^2+(y^2)^2+(y^3)^2
\label{Euclidean}
\end{equation}
as required. The hat operation in turn induces the following
conjugation on 2-spinors:
\begin{equation}
\omega^{A}=(a,b)\rightarrow\hat{\omega}^{A}=(-\bar{b},\bar{a}),\qquad\pi_{A'}=(c,d)\rightarrow\hat{\pi}_{A'}=(-\bar{d},\bar{c}),
\label{conjugation}
\end{equation}
such that this operation acts on twistors as follows:
\begin{equation}
Z^{\alpha}=(\omega^{A},\pi_{A'})\rightarrow\hat{Z}^{\alpha}=(\hat{\omega}^{A},\hat{\pi}_{A'}).
\end{equation}
Using this notation, the Dolbeault operator discussed in
section~\ref{sec:penrose_dolbeault} takes the form
\begin{equation}
  \bar{\partial}=d\hat{Z}^\alpha\frac{\partial}{\partial\hat{Z}^\alpha}.
  =d\hat{\omega}^A\frac{\partial}{\partial \hat{\omega}^A}
  +d\hat{\pi}_{A'}\frac{\partial}{\partial \hat{\pi}_{A'}}.
  \label{dolbeault_euclid}
\end{equation}
As explained in e.g. ref.~\cite{Adamo:2017qyl}, it is convenient to
rewrite this by introducing a particular basis for anti-holomorphic
vectors and (0,1) forms on the appropriate projective twistor space
(which we denote by $\mathbb{PT}(\mathbb{R}^4)$). That is, we may write
the tangent space of anti-holomorphic fields as
\begin{eqnarray} 
T^{0,1}_{\mathbb{PT}(\mathbb{R}^{4})}=\text{span}\left\{\bar{\partial}_{2}=\langle\pi\hat{\pi}\rangle\pi^{ A'}\frac{\partial}{\partial\hat{\pi}^{ A'}} ,\quad\bar{\partial}_{A}=\pi^{ A'}\frac{\partial}{\partial x^{ AA'}}\right\},
\label{vectors}
\end{eqnarray}
and the space of (0,1) forms as
\begin{eqnarray}
\Omega^{0,1}(\mathbb{PT}(\mathbb{R}^{4}))=\text{span}\left\{\bar{e}^{2}=\frac{\langle\hat{\pi}d\hat{\pi}\rangle}{\langle\pi\hat{\pi}\rangle^{2}},\quad\bar{e}^{A}=\frac{\hat{\pi}_{ A'}d x^{ AA'}}{\langle\pi\hat{\pi}\rangle}\right\},
\label{oneforms}
\end{eqnarray}
such that the Dolbeault operator of eq.~(\ref{dolbeault_euclid}) is
recast as
\begin{equation}
\bar{\partial}=\bar{e}^{2}\bar{\partial}_{2}+\bar{e}^{A}\bar{\partial}_{A}.  
  \label{dolbeault_euclid2}
\end{equation}
We may write the Penrose transform following eq.~(\ref{Penrose2}),
where the twistor function that appears must be an element of the
Dolbeault cohomology group
$H^{0,1}_{\bar{\partial}}(\mathbb{PT}(\mathbb{R}^{4}),\mathcal{O}(-2n-2))$,
and may be expanded in the above basis as
\begin{equation}
f=f_{2}\bar{e}^{2}+f_{A}\bar{e}^{A}.
\label{fexpand}
\end{equation}
The convenience of this basis then becomes apparent. Upon restriction
to the Riemann sphere $X$ corresponding to fixed $x^{AA'}$, only the
term involving $\bar{e}^2$ survives, and one thus finds
\begin{equation}
\phi_{A'B'\ldots C'}(x)=\frac{1}{2\pi i}\int_{X}\langle\pi d\pi\rangle\wedge\pi_{A'}\pi_{B'}\ldots \pi_{C'}f_{2}\vert_{X}\bar{e}^{2}.
\end{equation}
Let us now return to the problem of how to pick special
representatives of Dolbeault representatives for given spacetime
fields, such that the twistor double copy prescription of
eq.~(\ref{Fgrav}) becomes more meaningful. First, let us recall that
on a complex manifold $M$, one may define an positive definite inner
product between two $(p,q)$ forms $\alpha$, $\beta\in\Omega^{p,q}(M)$
according to
\begin{equation}
  (\alpha,\beta)\equiv\int_M\alpha\wedge \ast \bar{\beta}.
  \label{innerproduct}
\end{equation}
We may then define the {\it adjoint Dolbeault operators}
$\partial^\dag$, $\bar{\partial}^\dag$ via 
\begin{equation}
  (\alpha,\partial\beta)=(\partial^\dag\alpha,\beta),\quad
  (\alpha,\bar{\partial}\beta)
  =(\bar{\partial}^\dag \alpha,\beta).
  \label{adjoints}
\end{equation}
Then, the {\it Hodge decomposition theorem} says that, if $M$ is
compact, one may write an arbitrary $(p,q)$ form
$\omega\in\Omega^{p,q}(M)$ as
\begin{equation}
  \omega=\bar{\partial}\alpha+\bar{\partial}^\dag\beta + \gamma,
  \label{hodgedecomp}
\end{equation}
where $\alpha\in\Omega^{p,q-1}(M)$, $\beta\in\Omega^{p,q+1}(M)$ and
$\gamma\in\Omega^{p,q}(M)$. The form $\gamma$ is called the {\it
  harmonic part} of $\omega$ and satisfies
$\bar{\partial}\omega=\bar{\partial}^\dag\omega=0$. We denote by ${\rm
  Harm}^{p,q}_{\bar{\partial}}(M)$ the set of all such harmonic forms,
and there is a known isomorphism between the set ${\rm
  Harm}^{p,q}_{\bar{\partial}}(M)$ and the Dolbeault cohomology group
$H_{\bar{\partial}}^{0,1}(M)$, which is straightforward to understand:
elements of the latter are cohomology classes; each cohomology class
has a unique harmonic representative, namely an element of the
former. For a $\bar{\partial}$-closed form ($\bar{\partial}\omega=0)$,
one finds $\bar{\partial}\bar{\partial}^\dag\beta=0$. Consideration of
\begin{displaymath}
\langle\beta,\bar{\partial}\bar{\partial}^\dag\beta\rangle
=\langle\bar{\partial}^\dag\beta,\bar{\partial}^\dag\beta\rangle
\geq 0
\end{displaymath}
then reveals $\bar{\partial}^\dag\beta=0$. \\

In our present context, we are concerned with (0,1) forms on the
Riemann sphere $X$ corresponding to a given spacetime point $x$. As
mentioned above, after restriction to $X$ one has
$\bar{\partial}|_X\equiv \bar{e}^2\bar{\partial}_2$. The above comments imply
that a $\bar{\partial}$-closed (0,1) form can be written as
\begin{equation}
f|_{X}=\bar{e}^2\bar{\partial}_{2}g+f_{\Delta},
\label{fXdecomp}
\end{equation}
where $g$ is a function and 
\begin{displaymath}
f_\Delta\in{\rm Harm}_{\bar{\partial}}^{0,1}(\mathbb{CP}^1,
\mathcal{O}(-2n-2))
\end{displaymath}
is a $\bar{\partial}$-harmonic (0,1)-form on $X\simeq\mathbb{CP}^{1}$,
where we have also labelled the homogeneity required for a spin-$n$
field. From eq.~(\ref{fXdecomp}), picking a Dolbeault representative
for a given field to correspond to the purely harmonic part amounts to
imposing the requirement\footnote{Equation~(\ref{HarmonicGauge}) also
occurs when describing Yang-Mills gauge fields in twistor space (see
e.g. refs.~\cite{Woodhouse:1985id,Adamo:2019lor,Mason:2005zm,Boels:2006ir,Jiang:2008xw,Adamo:2017qyl}
for this and related work), when it is referred to as the {\it
  harmonic gauge} condition. Our context here is more general, given
that the twistor function being referred to may describe a spacetime
field in either scalar, gauge or gravity theory.}
\begin{equation}
\bar{\partial}^{\dagger}_{2}f|_{X}=0,
\label{HarmonicGauge}
\end{equation}
and the Penrose transform then assumes the form 
\begin{equation}
\phi_{A'B'\ldots C'}(x)=\frac{1}{2\pi i}\int_{X}\langle\pi d\pi\rangle\wedge\pi_{A'}\pi_{B'}\ldots \pi_{C'}f_{\Delta}(Z)\vert_{X}.
\label{Penrose3}
\end{equation}
To address the relationship with the Weyl double copy, it is
worthwhile pointing out that there is an explicit mechanism to
generate harmonic Dolbeault representatives in twistor
space~\cite{Woodhouse:1985id}. Given a spacetime spinor field
$\phi_{A'B'\ldots C'}(x)$, one may construct a twistor function on $X$
as follows:
\begin{equation}
\Phi_\phi=\frac{1}{\langle\pi\hat{\pi}\rangle^{2n+1}}\phi_{A'B'\ldots
  C'}(x)\hat{\pi}^{A'}\hat{\pi}^{B'}\ldots\hat{\pi}^{C'}.
\label{Phiphidef}
\end{equation}
One may then construct the (0,1) form 
\begin{equation}
f_\phi=\hat{\partial}\Phi_\phi=\frac{2n+1}{\langle \pi\hat{\pi}\rangle^{2n}}
\phi_{A'B'\ldots C'}\hat{\pi}^{A'}\pi^{B'}\ldots \pi^{C'}\bar{e}^2,
\label{fphidef}
\end{equation}
where we have introduced the operator
\begin{equation}
  \hat{\partial}\equiv d\hat{\pi}^{A'}\frac{\partial}{\partial \pi_A'},
  \label{partialhat}
\end{equation}
and used the basis of eq.~(\ref{oneforms}). Equation~(\ref{fphidef})
indeed turns out to be harmonic. Conversely, using $f_\phi$ in the
Penrose transform of eq.~(\ref{Penrose3}) reveals $\phi_{A'B'\ldots
  C'}$ to be the spacetime field associated with the twistor one-form
$f_{\phi}$. To see this, one may write the Penrose transform out in
full as
\begin{eqnarray}
\phi_{A'\ldots C'}(x)&=&\frac{1}{2\pi i}\int_{X}\frac{\langle\pi d\pi\rangle\wedge\langle\hat{\pi}d\hat{\pi}\rangle}{\langle\pi\hat{\pi}\rangle^{2}}\pi_{A'}\ldots \pi_{C'}\frac{2n+1}{\langle\pi\hat{\pi}\rangle^{2n}}\phi_{D'\ldots E'}(x)\hat{\pi}^{D'}\ldots\hat{\pi}^{E'}\nonumber\\
&=&\frac{2n+1}{2\pi i}\phi_{D'\ldots E'}(x)\int_{X}\omega\frac{\pi_{A'}\ldots \pi_{C'}\hat{\pi}^{D'}\ldots\hat{\pi}^{E'}}{\langle\pi\hat{\pi}\rangle^{2n}}\nonumber\\
&=&\phi_{D'\ldots E'}(x)\delta^{D'}_{(A'}\ldots\delta^{E'}_{C')},
\end{eqnarray}
where
\begin{equation}
\omega=\frac{\langle\pi d\pi\rangle\wedge\langle\hat{\pi}d\hat{\pi}\rangle}{\langle\pi\hat{\pi}\rangle^{2}}
\label{volumeform}
\end{equation}
is the volume form on $\mathbb{CP}^{1}$ and we have used the
identity\cite{Jiang:2008xw}
\begin{gather}
\frac{1}{2\pi i}\int_{X}\omega\frac{\pi_{A'}\ldots \pi_{C'}\hat{\pi}^{D'}\ldots\hat{\pi}^{E'}}{\langle\pi\hat{\pi}\rangle^{2n}}=\frac{1}{2n+1}\delta^{D'}_{(A'}\ldots\delta^{E'}_{C')}.
\end{gather}
Thus, $f_\phi$ is the harmonic Dolbeault representative for the field
$\phi_{A'B'\ldots C'}$. \\

Consider now a pair of EM fields $\phi^{(1)}_{A'B'}$,
$\phi^{(2)}_{A'B'}$, a scalar $\phi$ and a gravity field
$\phi^{G}_{A'B'C'D'}$ that enter the Weyl double copy of
eq.~(\ref{WeylDC}). From these, we may construct twistor functions
$\Phi^{(i)}$, $\Phi$ and $\Phi^G$ according to
eq.~(\ref{Phiphidef}). It is then straightforward to verify that
eq.~(\ref{WeylDC}) implies
\begin{equation}
\Phi^G=\frac{\Phi^{(1)}\,\Phi^{(2)}}{\Phi}.
\label{Phiprod}
\end{equation}
For the gravity solution, we thus obtain a harmonic Dolbeault
representative
\begin{equation}
f^G=\hat{\partial}\left(\frac{\Phi^{(1)}\,\Phi^{(2)}}{\Phi}\right).
\label{fprodeuclid}
\end{equation}
We therefore see that the spacetime Weyl double copy implies a simple
product structure in twistor space. Furthermore,
eqs.~(\ref{Phiphidef}, \ref{fphidef}, \ref{Phiprod}) imply that all
Dolbeault representatives occuring in the scalar, gauge and gravity
theories are harmonic. This is perhaps the cleanest twistorial
incarnation of the double copy that we have yet encountered. Each
cohomology class corresponding to a given spacetime field has a unique
and minimal representative, namely that (0,1) form which is
harmonic. We have seen that it is possible to combine harmonic (0,1)
forms in twistor space of homogeneity $-2$ and $-4$, in order to
obtain a (0,1) form of homogeneity $-6$ that is also harmonic. This
then fixes which gravity solution we are talking about upon performing
the double copy. \\

Note that eq.~(\ref{fprodeuclid}) bears a resemblance to
eq.~(\ref{Fgrav}), i.e. to our first Dolbeault double copy obtained as
a simple rewriting of the \u{C}ech approach. However, there are
important differences: the (0,1) form of eq.~(\ref{Fgrav}) is defined
locally, in a single coordinate patch, whereas that of
eq.~(\ref{fprodeuclid}) is defined globally, as well as involving a
different differential operator. We may of course apply
eq.~(\ref{Fgrav}) in Euclidean signature, and it is clear that the
Dolbeault representatives defined by eqs.~(\ref{Fgrav},
\ref{fprodeuclid}) will not be the same in general. Nevertheless, for
gravity fields which obey the Weyl double copy, the two differing
representatives correspond to the same spacetime gravity field if the
same scalar and gauge fields are chosen. To see this, note that we may
Penrose transform each (0,1) form appearing in eqs.~(\ref{fF},
\ref{Fgrav}) to obtain a spacetime field, which we may then plug into
eqs.~(\ref{Phiphidef}, \ref{fphidef}) to generate harmonic
representatives. If the gravity field obtained by Penrose transforming
eq.~(\ref{Fgrav}) obeys the Weyl double copy, we may plug it into
eq.~(\ref{Phiphidef}) to give a function satisfying
eq.~(\ref{Phiprod}), as noted above. Then eqs.~(\ref{Fgrav})
and~(\ref{fprodeuclid}) correspond to the same spacetime gravity
field, namely to the Weyl double copy of the scalar and gauge
fields. \\

The above discussion implies that there are at least two choices of
Dolbeault representatives such that a product structure in twistor
space leads to the Weyl double copy in position space: those defined
separately by eqs.~(\ref{Fgrav}, \ref{fprodeuclid}). Unlike the case
of harmonic representatives, however, eq.~(\ref{Fgrav}) does not
furnish us with a clear interpretation of which representatives we
must choose in order to make the double copy manifest, (i.e. the
choice of \u{C}ech representatives appears ambiguous). As we have
already mentioned above, a third choice has recently appeared in the
literature~\cite{Adamo:2021dfg}, inspired by previous
work~\cite{MasonTN}. The authors considered purely {\it radiative
  spacetimes}, namely those that are completely defined by
characteristic data at future null infinity $\mathcal{I}^+$. Each
point $x$ in Minkowski spacetime is associated with a spherical
surface $S_x^2$, corresponding to where the lightcone of null
geodesics at $x$ intersects $\mathcal{I}^+$. Each null geodesic at
$x$, however, corresponds to a point in projective twistor space
$\mathbb{PT}$, and the set of all such points forms the Riemann sphere
$X$ associated with $x$. There is then a well-defined map from the
sphere $S_x^2$ to $X$, such that characteristic data on $S_x^2$ can be
used to fix a Dolbeault representative on $X$ whose Penrose transform
leads to a given radiative spacetime field~\cite{MasonTN}. As argued
in ref.~\cite{Adamo:2021dfg}, this may be done consistently in scalar,
gauge and gravity theories such that a spacetime double copy is
obtained. On the face of it, this procedure appears to be different to
either of the procedures defined above for choosing Dolbeault
representatives in the twistorial double copy, especially given that
ref.~\cite{Adamo:2021dfg} discussed radiative spacetimes only.

\section{Discussion}
\label{sec:discuss}

In this paper, we have considered the classical double copy
(specifically the Weyl double copy of ref.~\cite{Luna:2018dpt}), and
how one may formulate this in twistor space. This was already
considered in refs.~\cite{White:2020sfn,Chacon:2021wbr}, which showed
that a certain product of twistor functions can be used to derive the
Weyl double copy in position space. However, this creates a puzzle, in
that one cannot ordinarily multiply twistor functions together in the
Penrose transform that turns twistor quantities into spacetime
fields. The twistorial quantities associated with any spacetime field
can be subjected to equivalence transformations that do not affect the
latter, such that they are representatives of cohomology classes. This
casts doubt on whether the double copy can be furnished with a
genuinely twistorial interpretation, or whether the twistor approach
acts merely as a useful book-keeping device, that can be used to
efficiently generate instances of the classical double
copy. Furthermore, refs.~\cite{White:2020sfn,Chacon:2021wbr} used the
language of \u{Cech} cohomology groups, and if the twistor picture
makes sense then it must also be possible to instead use the more
widely used framework of Dolbeault cohomology.\\

We have herein presented two methods for writing the twistor double
copy in the Dolbeault framework. In the first, one may use a
well-known procedure for turning \u{C}ech representatives into
Dolbeault counterparts, in order to recast the Weyl double copy in the
Dolbeault language. The product structure that is inherent in the
\u{C}ech approach then survives in the Dolbeault approach, for obvious
reasons. Whilst it is encouraging that this works, it still provides
no clue as to how one can somehow pick out special representatives of
each cohomology class entering the double copy, so that the procedure
becomes unambiguous. To remedy this, we presented a second Dolbeault
double copy, which relies on known techniques for treating Euclidean
signature spacetime fields~\cite{WoodhouseTN,Woodhouse:1985id}. In
this approach, the Weyl double copy in position space indeed picks out
special cohomology class representatives in twistor space, namely
those (0,1) forms that are harmonic. This is particularly appealing
given that harmonic forms are uniquely defined for each cohomology
class, and in some sense minimal. However, it follows from the first
approach presented here that choosing harmonic forms is not the only
way in which a product in twistor space leads to the same Weyl double
copy in position space. Furthermore, neither of the approaches
presented here is obviously equivalent to the arguments of
ref.~\cite{Adamo:2021dfg}, which used characteristic data at future
null infinity to fix particular representatives corresponding to
radiative spacetimes. \\

We can perhaps clarify matters by considering the original double copy
for scattering amplitudes. In that case, it is only in certain {\it
  generalised gauges} (consisting of a choice of gauge and / or field
redefinitions) that the double copy -- which has a manifest product
structure term-by-term in a graphical expansion of the scattering
amplitude -- is made manifest. It is possible to work in different
generalised gauges, but at the expense of losing the simple product
form of the double copy~\cite{Bern:2017yxu}. Something like this idea
occurs elsewhere in the double copy literature, with a further example
being the Kerr-Schild double copy of exact solutions of
ref.~\cite{Monteiro:2014cda}. In that case, the gravity solution must
be in a particular coordinate system in order that its single copy can
be taken, which is such that a simple product formula applies between
the scalar and gauge fields entering the double copy. Kerr-Schild
coordinates are sufficient for this purpose, although there may be
other coordinate systems that accomplish this. An approach for copying
spacetime fields in arbitrary gauges at linearised level has been
developed in the convolutional approach of
refs.~\cite{Anastasiou:2014qba,LopesCardoso:2018xes,Anastasiou:2018rdx,Luna:2020adi,Borsten:2020xbt,Borsten:2020zgj,Borsten:2021hua},
which makes clear the the product form of the double copy is not
manifest in general. Based on these remarks, we find it highly
plausible that the double copy in twistor space can be given a general
form, such that the product structure is made manifest only for
particular cohomology representatives. That more than one product
structure leads to the same position space double copy is not a
problem, as there may be more than one choice of representatives that
makes the product structure possible. However, it seems unlikely that
the product structure would be true in general, given that it is so
obviously incompatible with the equivalence transformations that
define each cohomology class. This leaves the mystery of how one can
choose cohomology representatives {\it a priori} so that the twistor
space product applies. We regard our second Dolbeault double copy as
particularly useful in this regard, given that it {\it uniquely} fixes
a representative for each of the fields (scalar, gauge and gravity)
entering the double copy.\\

Our above remarks are of course only speculative, and other
possibilities remain. For example, one may have different product
structures that are possible in twistor space (corresponding to
different ways of picking cohomology representatives), but such that
these correspond to {\it different} double copies in position
space. In such a case, one could formally define the notion of a
(non-unique) {\it double copy} in twistor space by giving (i) a method
for choosing cohomology representatives for scalar, gauge and gravity
fields; (ii) a product formula (or other map) for combining the chosen
representatives. It may then turn out to be the case that only one of
these definitions matches the original double copy for amplitudes, but
the remaining double copies may nevertheless be useful for
something. The relationship between the twistor double copy of
refs.~\cite{White:2020sfn,Chacon:2021wbr} and the amplitudes double
copy has been very recently addressed in ref.~\cite{Guevara:2021yud},
which showed that classical spacetime fields can be obtained as a
Penrose transform of scattering amplitudes which have been transformed
from momentum to twistor space. The known double copy for amplitudes
would then imply a twistor-space double copy, and exactly how this
relates to the ideas of this paper would be very interesting to
investigate further. Another possibility is that there is no genuine
double copy in twistor space at all, and that the results obtained
thus far in refs.~\cite{White:2020sfn,Chacon:2021wbr,Adamo:2021dfg}
are coincidental, and do not generalise further. Our present paper
gives us hope that this is far too pessimistic a conclusion, but also
tells us that further investigation is necessary.

%%%%%%%%%%%%%%%%%

\section*{Acknowledgments}

We are grateful to Lionel Mason for discussions. This work has been
supported by the UK Science and Technology Facilities Council (STFC)
Consolidated Grant ST/P000754/1 ``String theory, gauge theory and
duality'', and by the European Union Horizon 2020 research and
innovation programme under the Marie Sk\l{}odowska-Curie grant
agreement No. 764850 ``SAGEX''. EC is supported by the National
Council of Science and Technology (CONACYT). SN is supported by STFC
grant ST/T000686/1.

\bibliography{refs}
\end{document}